\documentclass{mem}
\usepackage{natbib}\usepackage{txfonts}\usepackage{balance}
\usepackage{graphicx}
\usepackage[a4paper,breaklinks,dvipdfm]{hyperref}
\idline{84}{1}
\usepackage{txfonts} 

\begin{document}
\def\teff{$T\rm_{eff }$}
\def\kms{$\mathrm {km s}^{-1}$}

\title{A stellar-mass BH in a transient, low luminosity ULX in M31?}

%A stellar-mass BH in a transient, low luminosity 
%ULX in M31?

   \subtitle{}

\author{
Fabio Pintore \inst{1,2} \and Paolo Esposito\inst{3} \and Sara Motta\inst{4} \and Luca Zampieri \inst{1} }
          
  \offprints{F. Pintore}

\institute{
INAF -- Osservatorio Astronomico di Padova, Padova, Italy;\\
\email{fabio.pintore@studenti.unipd.it}
\and
Dipartimento di Astronomia, Universitˆ di Padova, Padova, Italy;
\and
INAF -- Istituto di Astrofisica Spaziale e Fisica Cosmica - Milano, Milano, Italy;
\and 
ESA/European Space Astronomy Centre, Villanueva de la Ca\~{n}ada, Madrid, Spain;
}

\authorrunning{F.~Pintore}

\def\xmm {\emph{XMM-Newton}}
\def\cxo {\emph{Chandra}}
\def\swift {\emph{Swift}}
\def\frm {\emph{Fermi}}
\def\sax {\emph{BeppoSAX}}
\def\xte {\emph{RXTE}}
\def\rst {\emph{ROSAT}}
\def\asca {\emph{ASCA}}
\def\src {XMMU\,J004243.6+412519}
\def\flux {\mbox{erg cm$^{-2}$ s$^{-1}$}}
\def\lum {\mbox{erg s$^{-1}$}}
\def\nh {$N_{\rm H}$}

\titlerunning{XMMU\,J004243.6+412519 in M31}

\abstract{We present a multi-wavelenght study of the recently discovered Ultraluminous X-ray transient XMMUJ004243.6+412519 (ULX2 hereafter) in M31, based on Swift data and the 1.8-m Copernico Telescope in Asiago (Italy). Undetected until January 2012, the source suddenly showed a powerful X-ray emission with a luminosity of $10^{38}$ erg s$^{-1}$ (assuming a distance of 780 kpc). In the following weeks, its luminosity overcame $10^{39}$ erg s$^{-1}$, remaining fairly constant for at least 40 days and fading below $10^{38}$ erg s$^{-1}$ in the next 200 days. The spectrum can be well described by a single multi-color disk blackbody model which progressively softened during the decay (from
$kT= 0.9$ keV to 0.4 keV). No emission from ULX2 was detected down to 22 \textit{mag} in the optical band and to $23-24$ \textit{mag} in the near ultraviolet. We compare its properties with those of other known ULXs and Galactic black hole transients, finding more similarities with the latter.
}
\maketitle{}

\section{Introduction}
Ultraluminous X-ray sources (ULX; e.g. \citealt{fabbiano89}) are extragalactic, point-like, off-nuclear sources with isotropic luminosity higher than $\sim$$10^{39}$ \lum. %High-quality X-ray spectra have shown that many bright persistent sources can be well fitted by a Comptonization model plus a disc (e.g. \citealt*{stobbart06,roberts07,gladstone09}). 
The nature of ULXs remains still matter of debate: 
%although their spectral properties may suggest sub-Eddington accretion onto Intermediate Mass Black Hole (IMBHs; 100-10000 $M_{\odot}$, \citealt{colbert99}),
several lines of evidence point towards super-Eddington or slightly super-Eddington accretion onto stellar or massive BHs (e.g. \citealt*{feng11,zampieri09} and references therein).
%point towards super-Eddington accretion onto stellar mass BHs (see \citealt{feng11} for a 
%review).
Till now, transient ULXs are still poorly investigated \citep[e.g. see][]{kaur12,soria12,sivakoff08}. In 2012 January a new X-ray source with a luminosity of $\sim$$10^{38}$ \lum\ was discovered by \xmm\ in M31 (\src\ or ULX2 hereafter; \citealt*{henze12}). Seven days after its discovery, it reached a luminosity of $\sim$ $2\times10^{39}$ \lum, which made it the second most luminous ULX in M31 \citep*{hph12} and possibly giving evidence of hosting a stellar mass BH \citep{esposito13,middleton13}.

\section{Results}

\begin{figure}[]
\resizebox{\hsize}{!}{\includegraphics[clip=true,angle=-90]{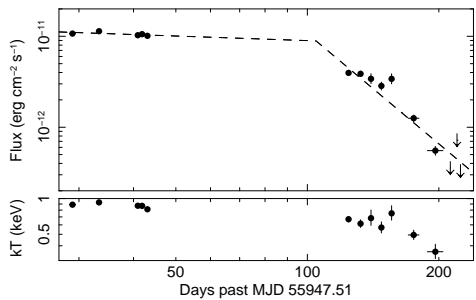}}
\caption{\footnotesize Top panel: time evolution of the absorbed flux in the 0.5-10 keV energy range for the disc component; the down-arrows indicate upper limits at the 3$\sigma$ confidence level. The broken-power-law model describing the decay is also plotted. Bottom panel: evolution of the characteristic temperature of the \textsc{diskbb} model inferred from the spectral fitting.}
\label{decay}
\end{figure}

ULX2 has shown a fast outburst, reaching luminosities higher than $10^{39}$ erg s$^{-1}$ and, after remaining almost constant for at least 40 days (we mention also an observational gap due to the Sun occultation), it subsequently decayed in the next 200 days. The lightcurve is described by a broken powerlaw (Figure~\ref{decay}-\textit{top}), assuming as t = 0 the date ULX2 was observed for the first time to exceed the ULX threshold. During the outburst and the following decay no short-term variability was detected and also nor QPOs up to 280 Hz.
\textit{Swift} spectra pre and post-gap may be well fitted by a single disc component, $\chi^2_{\nu}$(dof)$=$$1.07 (231)$ and $\chi^2_{\nu}$(dof)$=$$0.98 (66)$, respectively. The temperature of the disc decreases during the decay from $\sim0.9$ to 0.4 keV (Figure~\ref{decay}-\textit{bottom}).
No counterpart was detected in any of the UVOT observations and filters before or after the visibility gap and in the Asiago observations in the V and B band filters (down to a limiting magnitude of 21.7 and 22.2, respectively).
%UVOT filters - Pre-gap:
%u > 23.2 mag (total exposure: 13.4 ks),
%uvm2 > 23.7 mag (total exposure: 5.6 ks)
%uvw2 > 24.5 (total exposure: 18.2 ks)
%UVOT filters - Post-gap:
%uvw1 > 24.4 (total exposure: 23.2 ks

\section{Conclusions}
ULX2 does not show similarities with other known ULX transients. Its properties are reminiscent of those observed during the brightest state of many BH transients (e.g. XTE J1650Ð754, GROJ1655Ð4; \citealt{belloni11}). The X-ray spectrum, strongly dominated by a soft disc component, and the absence of short-time variability are consistent with the soft-state. We then suggest that ULX2 could be a stellar-mass BH binary. Assuming an accretion rate of 60$\%$ of the Eddington limit, the mass of the BH would be $\sim$12 $M_{\odot}$, consistent with that inferred by the normalization of the soft component, assuming that the inner disc radius is truncated at 6 gravitational radii and that the disc spectrum has a standard color correction factor (see \citealt{zampieri09,lorenzin09}). The optical/UV observations suggest a stellar mass BH accreting through Roche lobe overflow from a donor of main sequence star of 8-10 $M_{\odot}$ or a giant of $<$8 $M_{\odot}$ .
 
\begin{acknowledgements}
FP and LZ acknowledge financial support from INAF through grant PRIN-2011-1.
\end{acknowledgements}

\bibliographystyle{aa}

\end{document}